\documentstyle[aps,pra,twocolumn,epsfig,amsfonts]{revtex}

\renewcommand{\d}{\downarrow}
\renewcommand{\u}{\uparrow}
\newcommand{\s}{\sigma}
\newcommand{\p}{\partial_x}

\begin{document}

\title{Fermi 1D quantum gas: Luttinger liquid approach and spin-charge
separation}

\author{A. Recati,$^{1,2)}$ P.O. Fedichev,$^{1)}$ W. Zwerger$^{3)}$, P.
Zoller$^{1)}$}

\address{$^{1}$ Institute for Theoretical Physics, University of Innsbruck,
A--6020 Innsbruck, Austria\\
$^2$ Dipartimento di Fisica, Universit\`a di Trento and INFM, I-38050 Povo, 
Italy\\
$^{3}$ Sektion Physik, Universit\"at M\"unchen, Theresienstr. 37/III, D-80333 M\"unchen,
Germany.}

\maketitle
\begin{abstract}
We discuss the properties of quasi-1D quantum gases
of fermionic atoms using the Luttinger liquid theory,
including the presence of an optical lattice and of a 
longitudinal trapping potential. 
We analyze in particular the nature and
manifestations of spin-charge separation, where in the case of
atoms ``spin'' and ``charge'' refers to two internal atomic states
and the atomic mass density, respectively. 
\end{abstract}

\section{Introduction}

%Recently a renewed interest in quantum liquids with reduced dimensionality and their
%peculiar properties has appeared, expecially for what concern 1D systems. 
%This is mainly due to the recent experimental possibilities to trap ultra-cold 
%quamtum gases in quasi-1D magnetic and optical wave-guide. 
%Moreover, 
%as experimentally shown, one is able to include an underlying optical lattice. On which 
%seems turned the attention of many groups, both on the experimental and the theoretical
%side. 
The recent experimental possibilities to trap ultra-cold 
quantum gases in quasi-1D magnetic and optical wave-guide aroused 
a renewed interest in quantum liquids with reduced dimensionality
and their peculiar properties.
In this context, most of the recent theoretical and experimental work has
focused on quasi-1D Bose gases and Bose-Einstein condensates
\cite{salamon2002,1Dinteraction}. In particular, the equilibrium properties have 
been already discussed in several papers (see Ref.\cite{1DBose-equili} and 
references therein), the 
phase fluctuations of the classical field (order parameter) describing a quasi-1D 
condensate have been predicted in Ref.\cite{petrov-phase}, and experimental
evidence was reported in  Ref.\cite{hannover-phase}.
In Ref.\cite{ChiaraSS} the collective excitations for 1D Bose gases have been 
calculated for different
configurations, ranging from the 1D mean-field to the Tonks regime.   

At the same time also the possibility to include an optical lattice 
has become available, opening new perspectives in the study of coherence
phenomena \cite{Ioptlatt}.  
For instance Bloch oscillation have been seen in Ref.\cite{Morsch}. 
In Ref.\cite{JcurrentFI}, it is claimed that a direct measurement of the 
critical Josephson current has been obtained, and
very recently the intriguing quantum phase transition between superfluid and 
Mott-insulator, predicted some years ago in Ref.\cite{opticallattheory}, 
was observed 
for a Bose gas loaded in a 3D optical lattice in Ref.\cite{opticallatexp}. 
Also the dynamics of a trapped Bose-Einstein condensate in the presence of 2D and
1D optical lattice has been calculated in Ref.\cite{meretPRL}.

On the other hand remarkable progress in trapping and cooling Fermi gases, has been
made to reach the degenerate regime \cite{expfermi}. 
On the theoretical side different aspects of (harmonically) trapped Fermi gases have been explored, 
often in analogy with what has been done for Bose gases \cite{Rokhsar,Tosigroup,lorenzo}. 
Experimentally, measurements of the collective modes of degenerate Fermi gases 
have been reported (see Ref.\cite{expSpinExc} and reference therein). 

Fermi cold-gases, besides having an interest {\sl per se}, 
can represent a useful tool to implement solid state systems 
and to test fundamental theories, often grown 
and developed in a different area. 
Trapped gases are cleaner and more flexible systems 
than the solid state counterpart, where 
the experimentally most relevant 1D systems are semiconductor quantum wires and 
nanotubes. 
In addition, 
%it is a great advantage that 
it is also possible to use magnetic fields to largely modify the
scattering properties of atoms in atomic clouds\cite{Schreck}. 

%Most of the attention on quantum Fermi gases is 
%due on the possible experimental achievement of BCS-like phase transition. 
%However, as realized long time ago, one dimensional fermionic quantum liquids are very 
%rich and interesting system, also above BCS temperature. 
%For instance, just to cite one of the most , it is believed (P. W. Anderson) that typical 1D 
%properties can play an important role in the explanation of high-Tc superconductivity. 

Returning to 1D systems, what is interesting is that 
due to the reduced dimensionality the interaction, although small, 
strongly modifies the properties of the system. 
Low dimensions amplify the role of quantum fluctuations and enhance correlations. 
Both the ground state and the excitations
exhibit strong correlation effects and posses a number of exotic properties, e.g., 
fractional statistics  (see Ref.
\cite{HaldaneJPCPRL,Schulz:Review,lutt:fraqreview} and references therein).
Another fundamental interest in 1D systems originates from the fact that 
there is a fair number of exactly solvable models,
allowing the research of non-perturbative effects \cite{Tsvelik}. 
In particular Fermi-Landau theory which describes
3D interacting fermions is not longer applicable (e.g., see \cite{Schulz:Review}). 
Instead, in a quite wide range 
of parameters, 1D quantum liquids can be described using the Tomonaga-Luttinger model 
\cite{schoenhammer}, and in this case they are named {\sl Luttinger liquids}, in
contrast to (normal) Fermi liquids. 
The term Luttinger liquid was coined by Haldane \cite{HaldaneJPCPRL} to describe
the universal low-energy properties of gapless 1D quantum systems.    
The  Luttinger liquid language has been already applied to study the properties of 
(quasi)-1D Bose gases \cite{monien,Kagan}. 
%One of the hallmarks of a spin-$1/2$ Luttinger liquid (clearly the most interesting in
%the solid state context) is the so-called {\sl spin-charge separation}. 
%This feature involves the complete separation in the 
%dynamics of spin and charge degrees of freedom, both branches, which exhaust 
%(in a infinite system) the spectrum, are sound-like with different propagation velocities. 
%This phenomena has never seen in a clean way in actual condensed matter systems (see e.g.
%\cite{lutt:spinchargeobserve1}).    
The goal of this paper is to study trapped ultra-cold 
Fermi gases in quasi-1D geometry using Luttinger theory and to propose an
experimental way to see a typical Luttinger liquid behaviour, the so-called 
{\sl spin-charge separation}. Indeed one of the hallmarks of a spin-$1/2$ Luttinger 
liquid involves the complete separation in the 
dynamics of spin and charge degrees of freedom. 
Both branches, which exhaust 
(in a infinite system) the spectrum, are sound-like with different propagation 
velocities.
This phenomena has never seen in a clear way in actual condensed matter systems (see e.g.
\cite{lutt:spinchargeobserve1}).    

The paper, which can be considered a self-consistent, extended version of Ref. \cite{recati}, 
is structured as follows.
In SEC. \ref{sec:model} we introduce the models. 
Using standard argument, we write down a
bosonized Hamiltonian to describe the low-energy excitations of the systems 
(SEC. \ref{sec:ll}).
We used the available exact solutions to relate the so-called 
Luttinger parameters to the microscopic parameters of the system for small 
and large interaction (SEC. \ref{sec:ll}).
In SEC. \ref{sec:scsep} we shall propose different experimental ways to see 
clearly spin-charge separation in our systems, and in SEC. \ref{sec:relaxation} 
we study the life time of the excitations. 
In SEC. \ref{sec:bosons} we propose an implemetation of spinless Luttinger liquid using
strongly interacting bosons in an optical lattice. 

\section{Physical Model}\label{sec:model}

We consider a dilute gas of fermionic atoms of mass $m$ with two ground states,
$|\s=\u,\d\rangle$, representing a spin-$1/2$, e.g. the states
$|F=1/2, M_F=\pm 1/2\rangle$ of $^{6}$Li \cite{Li6}.
In the following we assume the number of atom in each level is the same: $N_\s=N/2$.
The atoms are confined in a very elongated trap along the $x$-direction
and cooled below the Fermi temperature $T_F$. The transverse confinement is 
considered to be harmonic with frequency $\omega_\perp$, strong enough so that 
the transverse degrees of freedom frozen. In order to have 
a (quasi) 1D system the tight transverse trapping must exceed the
characteristic energy scale of the longitudinal motion.
Due to the quantum degeneracy the longitudinal
motion has all the energy levels up to the Fermi-energy $\epsilon _{F}\sim k_{B}T_{F}$
occupied, thus the requirement is $\epsilon _{F}\ll \hbar \omega_\perp$. 

\begin{figure}
\centerline{\epsfig{file=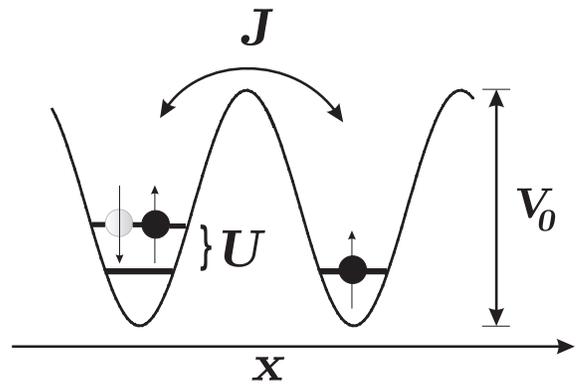,width=7.5cm}}
\vspace{0.3cm}
\caption{Schematic representation of fermions in an optical lattice.
The parameter $J$ is the amplitude to jump from one site to the adjacent one.
$U$ is the on-site energy due to the interaction between atoms in different
internal levels.}
\label{fig:lattice}
\end{figure}

At low temperature, due to the Pauli exclusion principle, only inter-species 
s-wave collisions are allowed. 
All the relevant interactions are thus characterized by a single 
parameter, the inter-component scattering length $a$. Introducing 
the harmonic oscillator transverse length $l_\perp =\sqrt{\hbar/m\omega_\perp}$, 
the effective 1D interaction can be represented as 
a zero-range potential with strength $g=2\pi \hbar ^{2}a/ml_{\perp }^{2}$, when  
$a\ll l_\perp$\cite{1Dinteraction}.
Thus the system can be described by the following 1D 
Hamiltonian 
%(as it is custom in literature, we also call Hamiltonian the 
%free energy operator)
\begin{eqnarray}
H& = &\sum_\sigma\int dx \psi_\sigma^\dagger(x)\left[\frac{-\hbar^2}{2m}\p^2
+V_{{\rm ext},\s}(x)\right]\psi_\sigma(x)\nonumber\\ 
& &+g\int dx
\psi_\u^\dagger(x)\psi_\d^\dagger(x)\psi_\u(x)\psi_\d(x),
\label{eq:H}
\end{eqnarray}
where $\psi_\s$ is the 1D field operator for atoms in the state
$\s$. The external potential $V_{{\rm ext},\s}$, includes both the longitudinal 
confinement $V_{L,\s}$ and a possible superimposed optical lattice 
$V_{{\rm opt},\s}$. 
The longitudinal confinement is either a box of length
$L$ (we will also refer to this as the homogeneous case), or an harmonic potential 
(inhomogeneous gas) with frequency $\omega_x$, i.e. $V_{L,\s}=1/2m\omega_x^2 x^2$. 
The
confinement potential can act differently on the two species, and this feature can 
be used in an experiment, as we will show below, to test the ``spin-charge
separation''.
The simplest way to create a 1D periodic potential for neutral atoms is by
superimposing two linearly polarized counter-propagating laser beams.
The created conservative potential is 
$V_{{\rm opt},\s}=V_0\sin^2(kx)$, FIG. (\ref{fig:lattice}), 
where the lattice period $d=\pi/k=\lambda/2$ is fixed by the wavelength, $\lambda$, of the laser
light (see, e.g., \cite{revMorsch}).  
%The presence of the lattice imply a band structure in the spectrum of the systems.
%We suppose that only states of the first band can be occupied, i.e, 
%we assume the system dynamics is small
%compared to the excitation energies of the second band. 
We can expand the field operators in the first band Wannier basis 
\begin{equation}
\psi_\s(x)=\sum_i w(x-x_i)c_{i,\s}, 
\end{equation}
where $w(x-x_i)$ is the Wannier function centered at the $i$-th site and 
$c_{i,\s}$ the annihilation operators for a fermions in the $i$-th site with spin $\s$.
Plugging the previous expression into Eq. (\ref{eq:H}) 
and retaining at most nearest neighboor terms for the kinetic energy part
we reduce the starting Hamiltonian to the Fermi-Hubbard Hamiltonian
\begin{eqnarray}
H_{FH}&=&-J\sum_{i,\s}(c^\dagger_{i+1,\s}c_{i,\s}+h.c.)+
\sum_{i,\s}\epsilon_{i,\s}n_{i,\s}\nonumber\\
& &+U\sum_in_{i,\u}n_{i,\d},
\label{eq:FH}
\end{eqnarray}
where operator $n_{i,\s}=c^\dagger_{i,\s}c_{i,\s}$ counts the number of 
$\s$-atoms in the site $i$. The parameter $U$ correspond to the strength of the on
site inter-species interaction depending on $g$, $J$ is the hopping term between adjacent sites
and $\epsilon_{i,\s}\simeq V_{L,\s}(x_i)$ describes an energy offset of each 
lattice site. Note that for $d\rightarrow 0$ one recovers a continuum model. 

\section{Luttinger liquid approach}\label{sec:ll}

In what follows we will often use the words charge and spin though in our system
we do not have, strictly speaking, charge nor spin. We used them to maintain the
jargon of the Luttinger physics. We identify the charge with the mass
and the spin with the two atomic internal levels. 
Thus when we talk about charge density
fluctuations we mean density fluctuations, when we talk about spin density
fluctuations we mean relative density fluctuations, the so called composite 
excitations.

\subsection{Homogeneous case}

In this subsection we will review briefly what it means to describe the low energy
properties of the system using the so-called Tomonaga-Luttinger model. 

We consider the homogeneous case, i.e., $V_{L,\s}=0$, and we will 
study later the inhomogeneous case. 
The standard way to represent the Eq. (\ref{eq:H}) for a gas in a box 
of length $L$ with periodic boundary condition,  without the underlying
optical lattice, in the momentum representation is
\begin{equation}
H=\sum_{k,\s}\varepsilon_kc^\dagger_{k,\s}c_{k,\s}
+\frac{g}{L}\sum_{k_1,k_2,q}c^\dagger_{k_1,\u}c_{k_1-q,\u}c^\dagger_{k_2,\d}c_{k_2+q,\d},
\label{eq:Hq}
\end{equation}  
where $\varepsilon_k=\hbar^2k^2/2m-\mu$. 

By means of renormalization group (RG), it is possible to show that the
low-energy properties of the system can be described using an exact solvable model,
the Tomonaga-Luttinger model \cite{HaldaneJPCPRL,Schulz:Review,lutt:fraqreview,schoenhammer}. 
First of all in a 1D system the Fermi ``surface'' is
composed by just two points, $\pm k_F$. In the non-interacting case one has 
$\pm k_F=\pm \pi \hbar n/2$, where $n=(N_\u+N_\d)/L$ is the total density of
the gas.
Let us, then, introduce the the left- (around $-k_F$) and right- 
(around $+k_F$) moving fermion annihilation 
(creation) operators $c_{r,\s}^{(\dagger)}(k)$, where $r=R,\; L$, and the respective
densities $\rho_{r,s}(q)=\sum_k c_{r,\s}^\dagger(k+q)c_{r,\s}(k)$.
In the model the four particle species $(R\u,R\d,L\u,L\d)$ have unbounded
free dispersion relation $\epsilon_{r,q}=a_r\hbar v_F q$, where 
$v_F=\hbar \pi n/2m$ is the Fermi velocity and $a_{R(L)}=+(-)$; 
all the states up to the Fermi energy, in the same spirit of the Dirac sea, are 
occupied. 
Thus one has two branches of particles with boundless energy and momentum and 
the densities obey Bose type commutation relation \cite{Schulz:Review}. 
The model Hamiltonian written in the notation of the so-called 
``$g$-ology'' (see for instance \cite{Solyom}), in real space representation, 
takes the form
\begin{eqnarray}
H_{LM}&=&\sum_\s\int dx \pi \hbar v_F(\rho_{R,\s}(x)^2+\rho_{L,\s}(x)^2)\nonumber \\
&+&g_{2\perp}\sum_\s\int dx \rho_{R,\s}(x)\rho_{L,-\s}(x)\nonumber\\
&+&g_{4\perp}\sum_{r,\s}\int dx\rho_{r,\s}(x)\rho_{r,-\s}(x)\nonumber\\
&+&g_{1\perp}\sum_{r,\s}\int dx
\psi^\dagger_{L,\sigma}\psi_{R,\sigma}\psi^\dagger_{R,-\sigma}\psi_{L,-\sigma}.
\label{eq:HLM_RG}
\end{eqnarray} 
Specific names are given to the parameters which identify the strength 
of a particular scattering process. For instance $g_{2\perp}$ is the strength of the
forward scattering between particles belonging to different branches and 
with different spin state, $(R\s,L-\s)\rightarrow(R\s,L-\s)$, while $g_{1\perp}$ 
is the strength of the backward scattering $(R\s,L-\s)\rightarrow(L\s,R-\s)$. In a
complete model also ``longitudinal'' scattering $(r\s,r'\s)\rightarrow(r\s,r'\s)$
could be present. 
In our case, where only inter-species collisions are allowed, we have  
$g_{2\parallel}=g_{4\parallel}=0$.
 
In the sense of RG in the long-wavelength limit, it is often
the case that $g_{1\perp}$ is renormalized to zero 
\cite{Schulz:Review,senechal,finite,fabrizio95}. This is true in our case.
The final step to solve the problem is to introduce four new boson fields
$\phi_\nu,\; \Pi_\nu$, with $\nu=c,s$, such that
\begin{equation}
\rho_{r,\s}(x)=\frac{1}{\sqrt{8\pi}}[\p\phi_c-a_r\Pi_c+\s(\p\phi_s-a_r\Pi_s)].
\label{eq:bosonfield} 
\end{equation} 
The bosonic fields are related to the fluctuations of the total density 
$\rho_c=\p\phi_c/\sqrt{\pi}$, of the spin-density $\rho_s=\p\phi_s/\sqrt{\pi}$, 
of the current density $j_c=-\Pi_c/\sqrt{\pi}$ and of the spin 
current density $j_s=-\Pi_s/\sqrt{\pi}$. The fields $\phi$ and $\Pi$ are conjugated
and they satisfies the bosonic commutation relation 
$[\phi_\nu(x),\phi_{\nu'}(x')]=[\Pi_\nu(x),\Pi_{\nu'}(x')]=0$ and 
$[\phi_\nu(x),\Pi_{\nu'}(x')]=i\delta_{\nu,\nu'}\delta(x-x')$.
Substituting Eq. (\ref{eq:bosonfield}) in Eq.
(\ref{eq:HLM_RG}) one obtains the bosonized Hamiltonian 
\cite{Schulz:Review}
\begin{equation}
H=\sum_{\nu=c,s} \frac{u_\nu}{2}\int dx \left[ K_\nu\Pi_\nu^2+\frac{1}{K_\nu}
(\p\phi_\nu)^2\right].      
\label{eq:HHaldane} 
\end{equation} 
We are left with the Hamiltonian of two independent elastic strings with the
eigenmodes corresponding to the collective density and spin-density fluctuation
of the fermion liquid. The parameters $u_\nu$'s are the ``sound'' velocities, 
while the $K_\nu$'s are related to the low energy behaviour of the correlation 
functions. In the following we will call them the Luttinger
parameters. They completely characterize the low energy 
physics. 

For excitations which do not change neither the number of particles and the currents,
the bosonic fields can be written in terms of boson creation
and annihilation operators, $b_{\nu ,k}^\dagger$ and $ b_{\nu ,k}$,
$\nu=c,s$
\begin{eqnarray}
\phi _{\nu }(x) & = & \sqrt{\frac{K_{\nu }}{2L}}\sum _{k}
\frac{1}{\sqrt{|k|}}\left( b_{\nu ,k}e^{ikx}+b_{\nu ,k}^{\dagger
}e^{-ikx}\right), 
\label{phi2q}
\end{eqnarray}
\begin{eqnarray}
\Pi _{\nu }(x) & = & i\sqrt{\frac{1}{2LK_{\nu }}}\sum _{k}
\sqrt{|k|}\left( b_{\nu ,k}e^{ikx}-b_{\nu ,k}^{\dagger }e^{-ikx}\right).
\label{Pi2q}
\end{eqnarray}

It is moreover possible to express the original fermionic fields in terms of the new
boson fields by the so-called bosonization identity
(see for instance \cite{Schulz:Review,senechal,vonDelft,Tsvelik}) 
\begin{equation}
\psi_{r,\s}=\lim _{\alpha \rightarrow 0}\frac{F_{r,\s}}{(2\pi\alpha )^{1/2}}
e^{-a_r i2\sqrt{\pi }\phi _{R(L),\s}}.
\label{eq:BI}
\end{equation} 
The fields $\phi_{r,\s}$ are recovered from Eq. (\ref{eq:bosonfield}).
The (hermitian) operators $F_{r,\s}$ are called Klein factors and obey the
Clifford algebra $\{F_{r,\s},F_{r',\s '}\}=2\delta_{r,r'}\delta_{\s, \s '}$.

%Let us now discuss a little more the Hamiltonian Eq. (\ref{eq:HHaldane}).

The same effective Hamiltonian, Eq. (\ref{eq:HHaldane}), can be obtained 
starting from the Hubbard Hamiltonian Eq. (\ref{eq:FH}). 
In the optical lattice case, besides the back-scattering term, also an umklapp 
contribution could in principle be present, but it does not contribute away from
half-filling. 

\subsubsection{Luttinger parameters}

For exactly solvable models like the 1D lattice model, the Luttinger parameters can 
be directly expressed in terms of the microscopic parameters of the theory 
\cite{Schulz:correxp}.
In a spin rotationally invariant Fermi-gas the quantity $K_{s}=1$, so
that the only independent parameters are $K_{\rho }$, and $v_{s,\rho }
$\cite{HaldaneJPCPRL,Schulz:Review}. 

One of the more peculiar consequences of the Hamiltonian Eq. (\ref{eq:HHaldane}) 
is a complete separation of the dynamics of the spin and charge degrees of freedom. 
In general one has $u_c\ne u_s$ and thus the density (or charge) and spin
oscillations propagate with different velocities. Only for a non-interacting gas, or 
for some accident, the two velocities are the same and equal to the Fermi velocity, 
$u_\nu=v_F$. This phenomenon is known as {\sl spin-charge separation}. 
Our aim is eventually to propose a way to see in a clear way such effect.\\
We should mention that if $g_{1\perp}$  renormalizes
to a finite value (in particular in a spin rotationally invariant system with bare 
$g_{1\perp}<0$, the parameter gets relevant by means of RG), 
one can have a gap in the spin sector, the charge excitations remaining massless 
\cite{Schulz:Review,senechal}.

In what follows we shall give some analytical results for the Luttinger parameters
of systems described by the Hamiltonian
Eq. (\ref{eq:H}) for a box and a harmonically trapping potential, 
with and without the underlying optical lattice.

For weak interaction, we define the small parameter $\xi=g/\pi \hbar v_F$.
We get, up to the first order in $\xi$, $u_c=v_F(1+\xi/2)$, $u_s=v_F(1-\xi/2)$, 
$K_c=1-\xi/2$. 
The same results are valid in the presence of optical lattice, the only difference 
being in the definition of $\xi=Ud/\pi \hbar v_F$, with Fermi velocity 
$v_F=2Jd\sin(\pi n d/2)/\hbar$ \cite{finite}.

When $V_{L,\s}=0$, i.e. in the homogeneous case, the models described 
by the Eq. (\ref{eq:H}) and Eq. (\ref{eq:FH}), are exactly solvable 
\cite{Yang,CollLieb}. 
The solutions confirm that the low energy spectrum is
Luttinger-like and so, at least in principle, it is possible to extract 
the Luttinger parameters for any interaction strength. 
%For small interaction, 
%to the first order in $\xi$, the parameters have the form we gave above. 
The exact solution \cite{Yang,CollLieb} is suitable to study the strong 
interacting limit, i.e., $\xi \gg 1$. 
The parameters can be found using perturbation theory around 
$\xi\rightarrow\infty$. 
Without the lattice one has: $u_s=2\hbar \pi n/3m\xi$, 
$K_c=1/2(1+8\ln(2)/\pi^2\xi)$ and $u_c=\pi \hbar n/m(1-8\ln(2)/\pi^2\xi)$, 
where $n=N/L$ is the density of the gas. 
When the lattice is present one obtains
\begin{eqnarray}
u_s=\frac{Jd}{\hbar\sin(\pi nd/2)\xi}\left(1-\frac{\sin(2\pi n d)}{2\pi nd}\right),\\ 
K_c=\frac{1}{2}\left(1+\frac{4\ln(2)}{\pi^2\xi}\frac{\sin(\pi n d)}{\sin(\pi n d/2)}\right),\\ 
u_c=\frac{2Jd\sin(\pi n d)}{\hbar}\left(1-\frac{4\ln(2)nd\cos(\pi
nd)}{\pi\sin(\pi nd/2)\xi}\right).
\end{eqnarray} 

Some remarks can be made. First of all we should mention that
the results without the lattice can be recovered from the previous expression 
expanding in the first (non-zero) order in low density $n$ and substituting 
$J\rightarrow\hbar^2/2md^2$, the new density being $n=N/N_w d\rightarrow N/L$,
with $N_w$ the number of sites. Indeed from the exact solution one gets that,
in the limit of low density, the lattice case resembles the continuous case 
with a renormalized mass $m^*=(2 J d^2/\hbar^2)^{-1}$.    
Secondly, note that the results (put $\xi\rightarrow\infty$) confirm the 
intuition that when the repulsion between the atoms of the two 
different species is very strong some properties of the gas are similar to 
those of an ideal single component gas of indistinguishable particles. 
In particular the Fermi velocity is
$v_F=\pi \hbar n/m$ and the spin velocity goes to zero. 
Intuitively one can say that, 
somehow, the infinite (hard-core or on-site) repulsion plays the role of an 
effective Pauli principle for atoms in different internal state 
(i.e., different spin). 
The same is true for instance for strongly interacting
bosons in 1D, where there exist an exact mapping between hard core bosons and 
free fermions \cite{girardeau}. 
Note, however, that in our case the
asymptotic value $K_c=1/2$\cite{Schulz:correxp}, while is $K_c=1$ for a 
non-interacting gas. 
It is well known that at zero temperature the Green's function of
a Luttinger liquids present a peculiar power-law decay with non-universal,
$K$-dependent exponents. At finite temperature the behaviour is still
$K$-dependent \cite{Schulz:Review}.
This means that physical properties like compressibility, density of states,
momentum distribution clearly reveal the interacting nature of the gas. 

\subsubsection{Inhomogeneous case: Local Density Approximation}

The inhomogeneous case will be studied using a local density approximation (LDA).
In order to apply LDA to our trapped gas we should assume that the size $R$ of the 
gas sample is much larger than the interparticle separation, i.e. $R\gg k_{F}^{-1}$. 
Such condition is consistent with having a large number of particle, $N\gg 1$. 
The ground state of the system
can be characterized using the Thomas-Fermi equilibrium condition:
\begin{equation}
\frac{dE_0(n)}{dn}=\mu -V_{L,\s}(x).
\label{eq:TF}
\end{equation}
Where $E_0(n)$ is the ground state internal energy of the gas per unit length, 
$n$ the total density and $\mu $ is the chemical potential. The longitudinal potential
in this case is $V_{L,\s}(x)=1/2m\omega_x^2 x^2$.
The previous equation is just the expression of the fact that the energy cost
of adding a particle to the system equals the chemical potential
corrected by the local value of the external potential.
In the Luttinger picture we assume that the spatial variation of 
$K_\nu$ and $u_\nu$ originate only from the spatial dependence of the gas density, i.e.,
$K_\nu[n]\rightarrow K_\nu[n(x)]$ and $u_\nu[n]\rightarrow u_\nu[n(x)]$.

The equation of motion for the fields conjugate fields representing the low energy 
excitations of the inhomogeneous gas are obviously given by 
$i\dot\phi_\nu = [H, \phi_\nu]$ and $i\dot\Pi_\nu = [H, \Pi_\nu]$, where $H$ is 
the Hamiltonian Eq. (\ref{eq:HHaldane}) with spatial dependent Luttinger paramters. 
Using the canonical commutation relation one can easily find:
\begin{equation}
\dot{\phi }_{\nu }=K_{\nu }(x)u_{\nu }(x)\Pi _{\nu },
\, \dot{\Pi }_{\nu }=\frac{\partial }{\partial x}\frac{u_{\nu }(x)}{K_{\nu }(x)}
\frac{\partial }{\partial x}\phi _{\nu }.
\label{eq:eqmot}
\end{equation}

Let us first of all consider a non-interacting fermion gas, i.e. $g=0$, without
the optical lattice.
This case is rather well-known (see for instance \cite{Rokhsar,Tosigroup}, where 
different approaches were used).
The density of the gas in the LDA in terms of the local value of 
Fermi momentum is $n(x)=2k_F(x)/\pi\hbar$.
The internal energy of the gas is just the density of the kinetic
energy (a.k.a. quantum pressure) $E_0(n)=\hbar ^2\pi ^2 n^3 (x)/24m$.
Substituting into Eq.(\ref{eq:TF}) we find the ground stete Thomas-Fermi density
profile
\begin{equation}
n_{TF}(x)=n_0\sqrt{1-\frac{x^2}{R_{TF}^2}},
\label{eq:nTF}
\end{equation}
for $|x|<R_{TF}$, and $0$ otherwise. 
We have indicated with $n_0=(8\mu m/\hbar ^2\pi ^2)^{1/2}$ the central density of the
trap and with $R_{TF}=(2\mu /m\omega_x^2)^{1/2}$ the Thomas-Fermi size of the cloud. 
The chemical potential $\mu =\hbar \omega_x N/2$ is fixed by the normalization 
requirement. 

For an ideal gas the equations of motion Eq. (\ref{eq:eqmot}) can be solved.
Indeed in this case the Luttinger parameters have the values $u_\nu=v_F$,
$K_c=1$, where the Fermi velocity $v_F$ can be written in terms of the total
density: $v_F=\hbar\pi n / 2m$. Substituting in the equation of motion Eq.
(\ref{eq:eqmot}) and using the density profile Eq. (\ref{eq:nTF}),
we can write down the equations for the density and spin-density fluctuations
($\propto \p \phi_\nu$): 
\begin{equation}
-\omega_\nu^{2}\rho _{\nu }=
\omega_x^{2}\frac{\partial}{\partial \tilde{x}}(1-\tilde{x}^2)^{1/2}
\frac{\partial}{\partial \tilde{x}}(1-\tilde{x}^2)^{1/2}\rho_{\nu },
\end{equation}
where we defined the adimensional coordinate $\tilde{x}=x/R_{TF}$. 
The same equation has been derived in \cite{Tosigroup} using the mean field 
equations of Kolomeisky et al.\cite{Kolomeisky}. The solution is given by 
\begin{eqnarray}
\sqrt{1-\tilde{x}^2}\rho_{\nu n}&=&A_\nu\sin 
(\omega_{\nu n}/\omega_x\arccos \tilde{x})\\
&+&B_\nu \cos (\omega_{\nu n}/\omega_x\arccos \tilde{x}).
\end{eqnarray}
By analyzing the boundary conditions one can find the discrete spectrum of 
eigenfrequencies. The result:  $\omega _{\nu n}=\omega_x n$ \cite{Tosigroup}
is the same both for the spin and the density modes. The first modes ($n=1$) has 
eigenfunction $\rho _\nu \sim x/\sqrt{1-\tilde{x}^{2}}$ and it corresponds to 
harmonic oscillations of the total density and the total spin 
(dipole and spin-dipole mode).

For weak interaction the correction to the ground state energy can be obtained 
by averaging the interparticle interaction over the free ground state:
\begin{equation}
E_0=\hbar ^{2}\pi ^{2}n^{3}/24m+gn^{2}/4.
\end{equation} 
Then, in the spirit of LDA using the Eq. (\ref{eq:TF}), one finds that the density, 
up to the first order in $\xi=g/\pi\hbar v_F(0)$,
uniformly decreases by $\delta n(x)=-2gm/\hbar^2\pi^2$. As expected the 
(repulsive) interaction reduces the density. 
This simple conclusion breaks down close to the edges of the cloud 
$x=\pm R_{TF}$ , where the density goes to zero. Indeed we should require 
that $g\ll \hbar^2 n(x)/m$, i.e., 
\begin{equation}
(R_{TF}-x)/R_{TF}\alt (gm/\hbar ^{2}n_{0})^{2}\sim O(\xi ^{2}).
\label{eq:border}
\end{equation}
The Luttinger parameters can be calculated substituting the local Fermi velocity 
$v_F(x)=\pi \hbar(n_{TF}(x)+\delta n(x))/2m$ in the expressions obtained 
before for the homogeneous case with small interaction. In this way one finds
\begin{equation}
u_{c,s}(x)=\frac{\pi \hbar n_{TF}(x)}{2m}\left(1-\frac{A_{c,s}gm}{\pi ^{2}\hbar
^{2}n_{TF}(x)}\right),
\label{eq:inhomvel}
\end{equation}
where $A_s=3$ and $A_c=1$ and to the same order $K_c=1-gm/\pi^2\hbar^2 n_{TF}(x)$. 

In the strong interacting limit from the exact solution one has 
\begin{equation}E_0=\hbar^2 \pi^2 n^3/6m(1 - 4\hbar^2 n \ln(2) /mg).\end{equation} 
The density profile for
$g\rightarrow\infty$ is still given by Eq. (\ref{eq:nTF}), but with 
$n_0\equiv n_\infty = (2m\mu/\pi^2\hbar^2)^{1/2}$, $\mu=N\hbar \omega_x$ and 
$R_{TF}=R_\infty = (2\mu/m\omega_x^2)^{1/2}$. The first order correction to the density is
$\delta n(x)=8\hbar^2\ln(2)n_{TF}(x)^2/3mg$. 
The spin velocity goes to zero as the interaction goes to infinity, the first order
correction is simply \begin{equation}u_s(x)=\pi^3 \hbar^3 n_{TF}(x)/3m^2 g.
\end{equation} 
The density wave
velocity and $K_c$ have a non-zero value for infinite interaction and up 
to the first order correction one has $K_c=1/2(1+4\hbar^2\ln(2)n_{TF}(x)/gm)$ and 
\[
u_c(x)=\frac{\pi \hbar
n_{TF}(x)}{m}\left(1+\frac{4\hbar^2\ln(2)n_{TF}(x)}{mg}\right).
\]

On the same lines we can obtain the results in presence of the optical
lattice. For the non-interacting case ($g=0\Rightarrow U=0$) the ground state 
energy of the gas is $E_0 = 2Jn-4J\sin(\pi n d/2)/\pi d$. From Eq. (\ref{eq:TF})
one has that the density of the cloud is 
\begin{equation}
n_{TF} = \frac{2}{\pi d}\arccos\left[1-\frac{\mu}{2J}\left(1-\tilde{x}^2
\right)\right]
\label{eq:lattnTF}
\end{equation}
for $\tilde{x}=x/R_{TF} \le 1$ and 0 elsewhere. From the previous expression we
find the equations of motion for
density and spin-density fluctuations in the ideal case
\begin{equation}
-\omega_\nu^{2}\rho _{\nu }=
\frac{m}{m^*}\omega_x^2\frac{\partial}{\partial
\tilde{x}}(1-\tilde{x}^2)^{1/2}f_l(\tilde {x})
\frac{\partial}{\partial \tilde{x}}(1-\tilde{x}^2)^{1/2}f_l(\tilde {x})\rho_{\nu },
\end{equation}
where we defined $f_l(\tilde{x})=\sqrt{1-\mu/4J (1-\tilde{x}^2)}$.
For low density (i.e. $f_l\rightarrow 1$) we have that the spectrum has the 
same structure as for the continuos case, but with a renormalized trapping frequency
$\omega_{\nu,n}=n\omega_x\sqrt{m/m^*}$. This result is quite general and for
instance the same correction was found in \cite{meret:1Doptlatt} for a
Bose-Einstein condensate in 1D optical lattice.

The first order correction, calculated averaging the interactionhn term in Eq.
(\ref{eq:FH}) 
over the ground state is $\delta n = -Un_{TF}/2\pi J \sin(\pi n_{TF} d/2)$. 
The density and spin-density wave velocities up to the first order are given 
by
\begin{equation}
u_{c(s)}=v_F(x)\left[1-\frac{Ud}{2\hbar v_F}
\left(\frac{n_{TF}d \cos(\pi n_{TF} d/2)}{\sin(\pi
n_{TF} d/2)}\pm\frac{1}{\pi}\right)\right]
\end{equation}
and to the same order the remaining parameter $K_c$ has the same expression than
in the homogeneous case but with a space dependent density given by the Eq.
(\ref{eq:lattnTF}).
In the strong interacting limit the energy density of the ground state can be written as
\begin{eqnarray}
E_0&=&2J\left(n-\frac{\sin(\pi n d)}{\pi d} \right)\nonumber\\
&+&\frac{2J^2\ln(2)}{Ud}\left[\frac{nd}{\pi}\sin(2 \pi n d)-(nd)^2\right].
\end{eqnarray}
The density profile for $U\rightarrow\infty$ is given by the Eq.(\ref{eq:lattnTF}),
divided by 2. 
The rest of the calculation proceeds on the same line as above. We
will use these results later to discuss spin-charge separation.

We have shown how it is possible, under certain condition, to study as a
Luttinger liquid an inhomogeneous 1D Fermi ultra-cold gas. 
We gave some analytical results in the two limiting case of weak and strong
interaction. This will be used in the 
next section to study a possible way to see the equivalent of spin-charge separation 
in such system.

\section{Spin-charge separation in trapped cold-gases}\label{sec:scsep}

\subsection{Spectroscopy}

The separation of the spin and the charge modes in the spectrum 
suggests experimental ways to observe the spin charge separation
in harmonically trapped atoms \cite{recati}. 
We will propose two related experiments to observe the separation of the spin
and the charge degrees of freedom.
One relies on the possibility to excite easily some normal modes of the trapped gas
and to measure their frequency.
The other one consist in really testing the dynamics of total density and relative
density wave packets.
Before discussing the experiments it should be noticed that the application of the
Hamiltonian Eq. (\ref{eq:HHaldane}) to finite systems deserves special attention.
As shown in \cite{finite} the boundaries of a LL affect the physics of the lowest
energy excitation $\epsilon$ with $\epsilon L/v_F\leq 1$. 
Setting $L\sim R_{TF}$ we find
that the excitations with $\omega_{\nu n}\sim \omega_x$ can have finite size
corrections, which are not described by Hamiltonian Eq. (\ref{eq:HHaldane}) with the
LDA values of the Luttinger parameters. Below we will consider $\omega_{\nu
n}\gg\omega_x$ (WKB approximation, see below) or small interactions. Hence we can
neglect the finite size effect. Other finite size effect related to the TF
approximation will appear (logarithmic correction), but they are not related to
Luttinger physics.

\begin{figure}
\centerline{\epsfig{file=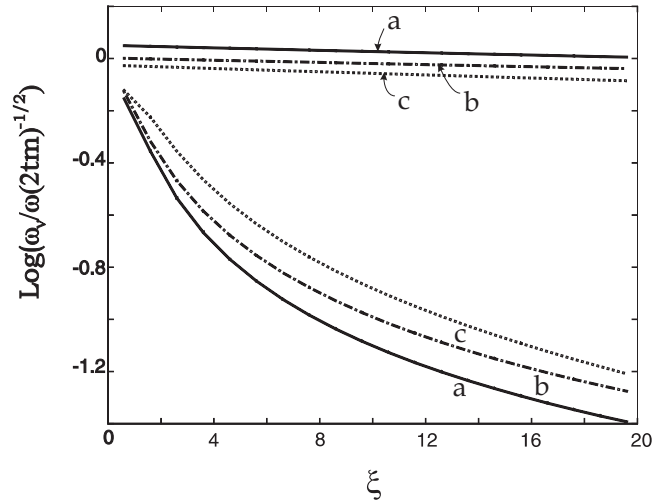,width=8.5cm}}
\vspace{0.3cm}
\caption{Level spacing (in units of $\omega_x$) of the spin
(lower curves) and the charge (upper curves) modes for a harmonically trapped Fermi gas 
in the presence of an optical 
lattice (see the text). The level spacing is shown as a function of $\xi=g/\pi \hbar v_F(0)$ 
and for different central densities:\\
curves a: $n(0)=0.25/d$ (solid lines), \\
curves b: $n(0)=0.42/d$ (dot-dashed lines), \\
curves c: $n(0)=0.58/d$ (dotted lines).}
\label{fig:frequencies}
\end{figure}

Assume we have the two different components
of an interacting Fermi-system confined in external potentials, which
act independently on each component. Then by modulating the two potentials
simultaneously one induces oscillations of the total density. In turn,
by slightly changing the confining potentials for the two components
out of phase, only spin waves are excited. By observing the character
and measuring the frequencies of the modes one can prove that the
spin and the density modes are in fact the independent modes in the
system and test the validity of the Luttinger description for such system. 
In FIG. \ref{fig:frequencies} we show the numerical results of the level spacing
for the spin and the charge modes for harmonically trapped atoms in the presence of optical lattice.
Numerical results are based on the exact solution \cite{CollLieb} together with
the LDA.

For small interaction we can give an analytical expression to the energy 
correction of the lowest modes for a trapped gas using the balance equation Eq.
(\ref{eq:TF}). The frequency of the (density) dipole mode is unchanged as it
has to be, while the correction to the spin dipole mode diverge
logarithmically. The logarithmically divergent behaviour is due to the localization of the
excitations of a free gas close to the border (see before and
\cite{Tosigroup}). To handle the divergence we use as cut-off the
condition Eq. (\ref{eq:border}) and we find
\[
\delta \omega _s=-\omega_x \frac{3gm}{\pi ^{2}\hbar ^{2}n_{TF}(0)}
\log \frac{\pi \hbar ^{2}n_{TF}(0)}{2gm}.
\]
The energy shift is negative and can be measured by comparing the density and
the spin oscillation of the atomic cloud. The logarithm in the previous
expression tells us that in harmonic traps the requirement to apply the
perturbation theory in $\xi$ is stronger than in the homogeneous case, namely
it should be $\xi\log(1/\xi)\ll 1$. We can say that the LDA is valid with logarithmic
accuracy.

The same behaviour for the first order 
correction to the spin-dipole frequency can be obtained in this regime 
also using sum rules. This was done for a 3D system in \cite{lorenzo}.
Generally speaking, it is possible to show, that an upper bound to the energy 
$\hbar\omega_F$ of the lowest state excited by an operator $F$ is given by
the ratio $\hbar^2\omega_F^2=m_3/m_1$, where $m_k$ is the $k$th
moment of the dynamic structure factor. One can write the two moments as 
\begin{eqnarray}
m_1&=&\frac{1}{2}\langle0|[F^\dagger,[H,F]]|0\rangle, \nonumber \\
m_3&=&\frac{1}{2}\langle0|[[F^\dagger,H],[H,[H,F]]]|0\rangle,
\end{eqnarray}
where $H$ is the Hamiltonian
and $|0\rangle$ the ground state of the system. We are interested in the
excitations due to the spin dipole operator $S=\sum_{i\u}x_i-\sum_{i\d}x_i$ and the
Hamiltonian being Eq. (\ref{eq:H}).
In the limit of small interaction one has 
\begin{equation}
\omega_s=\omega_x-\frac{g}{mN\omega_x}\int dx |\p n(x)|^2.
\label{eq:srules}
\end{equation}
Substituting for the ground state density $n$, the density profile Eq.
(\ref{eq:nTF}) and using the cut-off Eq. (\ref{eq:border}), 
one finds, to logarithmic precision
\begin{equation}
\delta\omega_s=-\omega_x\frac{16}{\pi}\frac{gm}{\pi^2\hbar^2 n_{TF}(0)}
\log\left(\frac{\hbar^2 n_{TF}(0)}{gm}\right).
\end{equation}

The Eq. (\ref{eq:srules}) has nothing to do with Luttinger model and, hence, provides
an alternatives derivation for the spin excitation frequency. The fact that the
results agree with each other confirms both the applicability of the Luttinger model and 
the LDA. 

We will introduce now a WKB calculation of the energy excitation, which we will allow
us to calculate the energy shift of the spin excitations in the strong interacting
regime.
Suppose that the energy levels of a quantum system are charcterized by a certain 
quantum number $n$. For sufficently high quantum number the 
eigenfrequencies can be obtained from the WKB (Bohr-Sommerfeld) quantization 
condition,
\begin{equation}
\int _{-x_{0}}^{x_{0}}p(x)dx=\hbar \pi (n+\alpha ),
\label{bornsommer}
\end{equation}
where $p(x)$ is the WKB momentum corresponding to a given energy,
$x_{0}$ is the classical turning point and the constant $\alpha$ is the so called
Maslov index. 
The accuracy of the WKB spectrum estimation is $\sim 1/(\pi n)^2$.
In our case we can fix $\alpha$ by comparing the WKB results and the
exact solutions of Eqs. (\ref{eq:eqmot}) for a weakly interacting
gas. 
The spectrum of the excitations we are considering is linear, 
$\epsilon =u_{\rho ,s}(x)p(x)$,
with the velocities given by the Eq. (\ref{eq:inhomvel}). Using the 
Eq. (\ref{bornsommer}) we obtain the same sort of logarithmically diverging integrals
as those we obtained in the perturbative calculation above. Once again to regularize 
such diverging behavior we use the condition (\ref{eq:border}) and we find
\begin{equation}
\epsilon _{\rho ,s}=\hbar \omega (n+1)
\left(1-\frac{2gmA_{\rho ,s}}{\pi ^{2}\hbar ^{2}n_{TF}(0)}
\log \left(\frac{\hbar ^{2}n_{TF}(0)\pi }{mg}\right)\right).
\label{eq:spectrum}
\end{equation}
We find once more that the interaction dependent splitting of the eigenenergies
is different for charge and spin.

In the strong interaction limit, after integrating Eq. (\ref{bornsommer}), 
we find that once again the density dipole mode is unchanged. 
The leading term of the energy of the spin dipole mode, which is
zero in the limit $\xi\rightarrow\infty$, is given by 
\[
\epsilon _{s}(p)=\frac{\hbar^4 \pi^3 n^2(x)p(x)}{3 m^2 g}.
\]
Substituting $p(x)$ in the integral equality (\ref{bornsommer}) and dealing 
with the logarithmic divergence as usual, we find, that
\begin{equation}
\epsilon _{ns}=\hbar \omega (n+\alpha )\frac{\hbar ^{5}\pi^3 n_{\infty }}{3gm\log (gm/\hbar ^{2}n_{\infty })},
\end{equation}
where $\alpha \sim 1$. 
In the limit of the strong interaction the spin level spacing decreases a lot
and is much smaller than that between the density waves ($\omega_x$), as is clearly
shown in FIG.\ref{fig:frequencies}.

\subsection{Wavepacket dynamics}

\begin{figure}
\centerline{\epsfig{file=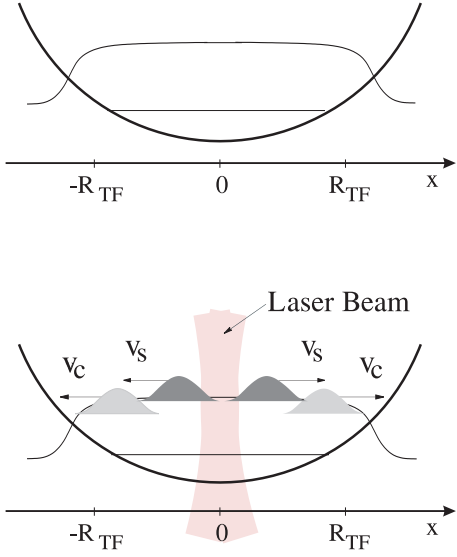,width=6cm}}
\vspace{0.3cm}
\caption{Two component quasi-1D Fermi gas in a harmonic trap. A short laser
pulse focused near the center of the trap can excites density end spin density
wave packet.}
\label{fig:trap-wp}
\end{figure}

\begin{figure}
\centerline{\epsfig{file=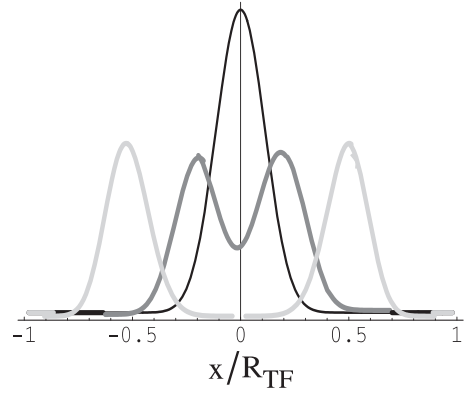,width=6cm}}
\vspace{0.3cm}
\caption{Wave packet dynamics for different times as function of position 
(in units of the Thomas Fermi radius $R_{TF}$): spin-charge separation
manifests itself in a spatial separation of the spin (solid line)
and density (dashed line) wave packets (shown at half a trap
oscillation period $\omega t=\pi/2$), which can be probed by a
second short laser pulse at a later time. The parameters
correspond to $N=10^3$ $^6$Li atoms in a trap with $\omega=1$Hz
(with coupling parameter $\xi=1$, see text).} 
\label{fig:wp-dyn}
\end{figure}

The difference between the spin and the charge velocities allows
to have spin and charge wave packets moving at different velocities. 
This can be seen in a sufficiently long trap. Localized wave packets of the 
spin and density excitations can be generated by sufficiently localized optical
potentials of dimension $\ell $ such that $R\gg \ell \gg k_F^{-1}$, where $R$
is the size of the atom cloud and the Fermi momentum $k_F$ is essentially the
interparticle distance FIG. \ref{fig:trap-wp}. 
The procedure we have in mind is
analogous to the MIT setup originally used to study propagation of
sound waves in elongated condensates \cite{MIT:soundprop}. The motion of these wave 
packets is readily
detected by internal state dependent laser probes. 
In FIG. \ref{fig:wp-dyn}
we show a simulate wave packets motion for the states
$|F=1/2, M_F=\pm 1/2\rangle$ of $^{6}$Li with interaction
parameter $\xi=1$. This figure was obtained solving numerically Eq. (\ref{eq:eqmot}).
The scattering length of these two states is magnetically tunable and recently the
tunability has been experimentally explored up to high magnetic field 
(see \cite{IBKexp2002} references therein). 
The scattering length is positive for applied magnetic field
$B\ge 550G$ and presents a Feshbach resonance around $800G$. 
We can, for instance, consider small scattering length of the order of $a= 20 \AA$. 
In order to have $\xi\sim 1$, one can use $N=500$ particles at trap
frequencies $\omega_x=1$Hz and $\omega_\perp=250$kHz. 
On the other hand near the Feshbach resonance the scattering length can be larger 
by orders magnitude, with respect to the previous value. For $a$ just ten times 
bigger than the previous one it is possible to have $N=1000$ atoms at a trap frequency 
$\omega_\perp=100$kHz, leaving unchanged $\omega_x$.

\begin{figure}
\centerline{\epsfig{file=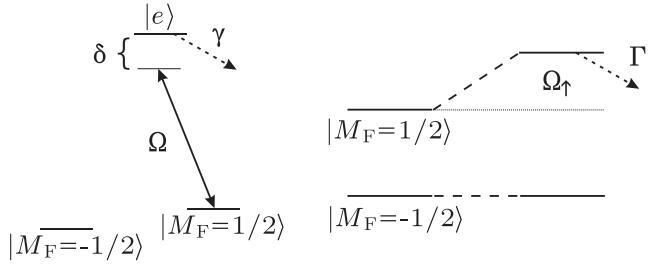,width=8.5cm}}
\vspace{0.3cm}
\caption{Internal level scheme of the atoms of the gas. The state 
$|\u\rangle=|M_F=1/2\rangle$
is coupled via a laser to a (in general unstable) state $|e\rangle$ (left). The
coupling being given by the Rabi frequency $\Omega$. 
On the right, we sketched, the effective action of the laser for large 
detuning $\delta\gg\Omega,\gamma$ (see the text).} 
\label{fig:laser}
\end{figure}

An optical potential created by a far detuned
laser light and acting in the same way on both the components
amounts to an external potential $\sim \p\phi_c$ and
hence acts only on the density waves. An optical potential generated
by a laser tuned, e.g., between fine structure levels of excited Alkali
states acts on the ground state spin in a way equivalent to
an external magnetic field interacting with the spin density $\sim\p\phi_s$
and thus act solely on the spin waves. 
One can also act at the same time coherently on both the charge and the spin waves.
We analyze in some detail this case (the same can be done for the previous cases).
The idea is sketched in FIG. (\ref{fig:laser}). 
We need to couple only one out of the two 
(``long-living'') hyperfine levels identified by $|\s\rangle = |\u\rangle, |\d\rangle$,
with another (in principle ``short-living'') state of the atom, $|e\rangle$. 
Let $\Omega$ be the Rabi frequency which couples 
for instance $|\u\rangle$ and $|e\rangle$, $\delta$ be the detuning, i.e. 
$\delta=\omega_L - \omega_{\u e}$, where $\omega_L$ is the laser frequency and
$\hbar\omega_{\u e}$ is the energy separation between $|\u\rangle$ and $|e\rangle$, and  
$\gamma$ be the spontaneous emission rate of the state $|e\rangle$. 
In the regime of large detuning, $\delta\gg \Omega,\gamma$ the effect of the coupling 
is to introduce an energy shift $\Omega_\u=|\Omega|^2/\delta$ and a decay rate
$\Gamma=|\Omega/\delta|^2\gamma$. Thus, for not too long times ($\Gamma t \ll 1$), 
we have the Hamiltonian density 
${\mathcal H}_{\rm ext}(x)=\Omega_\u(x)\psi_\u^\dagger(x)\psi_\u(x)$ 
acting on the gas. Eventually, if $\Omega_\u(x)$ (i.e. the laser profile) is slowly
varying in space, to avoid scattering processes,  $q\simeq 2k_F$, 
in terms of charge and spin fields one can write
\begin{equation}
H_{\rm ext}=\int\frac{dx}{\sqrt{2\pi}}\Omega_\u(x)(\p\phi_c(x)+\p\phi_s(x)).
\end{equation}
Clearly the previous Hamiltonian acts as localized (with the above remark) independent 
perturbations on both charge and spin densities.

\section{Relaxation phenomena}\label{sec:relaxation}

As stated before, the Hamiltonian Eq. (\ref{eq:HHaldane}) describes properly the
fluctuations of a 1D system only in the long wave $q/k_F\ll 1$, low energy regime. 
Higher order terms originating from, for example, the non-linearity of the fermionic 
spectrum appear which can mix the excitations. The first corrections
are third power in $\Pi_\nu$ and $\phi_\nu$, which lead to scattering of the
excitations.  
This is precisely the case in the simple model described by Eq. (\ref{eq:Hq}). 
Introducing once again the operators $c_{r,k,\s}$, $r=R,L$ and $\s=\u,\d$ 
for the left- and right-movers in the momentum space, but this time 
keeping the terms up to the second order in $k-k_F$, one has for the kinetic part
\begin{eqnarray}
H_0&=&\sum _{r,k,\s}(\partial_k\epsilon)_{k_F}(a_r k-k_F)c_{r,k,\s}^\dagger
c_{r,k,\s}\\
&+&\frac{(\partial^2_k\epsilon)_{k_F}}{2}(a_r k-k_F)^2c_{r,k,\s}^\dagger
c_{a,k,\sigma },
\end{eqnarray}
where $a_{R(L)}=+(-)$. 
The quadratic term in real space reads 
$(\partial^2_k\epsilon)_{k_F}\sum _{r,\s}\int dx:\p\psi
_{r,\s}^\dagger(x)\p\psi _{r,\s}(x)$ and using the bosonization identity (see
Appendix Eq. (\ref{eq:BI})) one obtains  
$4\sqrt{\pi}(\partial^2_k\epsilon)_{k_F}\sum _{r,\s}\int dx(\p\phi _{a,\s})^3$.
Introducing the charge and spin boson fields Eq. (\ref{eq:bosonfield}),
we find that we have to add the following third order term to the
Luttinger model density Hamiltonian
\begin{eqnarray}
{\mathcal H}_{\rm int}&=&\frac{(\partial^2_k\epsilon)_{k_F}}{\sqrt{2}\pi}\{(\p\phi_c)^3+3\Pi
_c(\p\phi_s\Pi_s+\Pi_s\p\phi_s)\nonumber\\
&+&\p\phi_c\Pi _c^2+\Pi_c\p\phi_c\Pi_c+\Pi_c^2\p\phi_c\nonumber \\
&+&3\p\phi_c[(\p\phi_s)^2+\Pi_s^2]\} .
\end{eqnarray} 

To study the relaxation phenomena we will consider, for simplicity, a single 
component gas (polarized system). 
In the previous Hamiltonian we are left only with the
term ${\mathcal H}_{\rm int}=\hbar^2 (\p\phi)^3/\sqrt{2}\pi m$. \\

According to Haldane \cite{HaldaneJPCPRL}, the generic term due to the non-linearity 
of the spectrum is given by $V_{\rm int}=\gamma\hbar^2/m\int(\p\phi)^3$, 
where $\gamma\sim 1$.

We calculated the damping of the oscillations by the Fermi's Golden rule and the
second quantized representation for the phonon fields Eqs. (\ref{phi2q}) and 
(\ref{Pi2q}). 
We considered the decay rate of an excitation with momentum $q$ and energy $\omega_q$.
In the limiting case $k_{B}T\ll \hbar \omega _{q}$, a straightforward calculation 
gives \begin{equation}\Gamma _{T=0}\sim \hbar ^{2}q^{4}L/m^{2}u.\end{equation} 
This is the decay rate in a process
where a particle with the energy $\hbar \omega _{q}$ decays into
a pair of particles with $\omega _{1},\omega _{2}<\omega _q$.
Note that $\omega _{q}\gg \omega_x $, otherwise there are no final
states for such decay instability (i.e. the lowest excitations are
very stable at very low temperatures). In the case of $k_{B}T\gg \hbar \omega _{q}$ 
(but still $T\ll T_{F}$) we find \begin{equation}\Gamma _{T}\sim
q^{2}Lk_{B}^{2}T^{2}/m^{2}u^{3}\end{equation}
which corresponds to Landau damping, i.e., the contribution of a process
in which the damping occurs by scattering a high-frequency excitation
with $\hbar \omega _{q}^{\prime }\sim k_{B}T$. 
Note that both results contain the size of the sample $L$. 
For sufficiently small temperatures (or high number of particles) 
the excitations are only weakly damped. Indeed, for the
lowest excitations $Lmu/\hbar \sim N$, 
$\omega _{q}\sim \omega \sim \epsilon _{F}/\hbar N$,
thus substituting the previous expression for the decay rate we have
\begin{equation}
\Gamma /\omega _{q}\sim (k_{B}T/\epsilon _{F})^{2}\ll 1.
\end{equation}

In conclusion for sufficiently low tempertaure or large enough number 
of particles the excitations are only weakly damped by the third order correction 
to the otherwise quadratic Luttinger Hamiltonian. 
This means that in a degenerate quasi-1D Fermi gas
the excitations can be observed for several trap periods.

\section{Spinless LL with bosons}\label{sec:bosons}

In this section we will show how to realize a spinless LL Hamiltonian
using ``hard core'' bosons in an optical lattice. 
The on site interaction of the bosons can be written as $U(n_{i}-1)n_{i}$. 
Hence, in the limit of $g\rightarrow \infty $ $(U\rightarrow \infty)$ 
only the states $n_{i}=0,1$
are allowed. We parameterize them as $\alpha =\pm $ states of a particle
with the spin $1/2$.  In terms of spin operators (Pauli matrices) 
the lattice (Hubbard) Hamiltonian can be put in the form 
\begin{equation}
H=J\sum _{i}(\s_i^+\s_{i+1}^- +h.c.)+V\sum (\s_i^z+1)(\s_{i+1}^z+1).
\label{eq:Hspin}
\end{equation}
Where $V$ is due to the off-site interaction, which is present in the starting
bosonic Hamiltonian. Usually one has $V\ll U$. In the easiest case, if $d$
is the lattice period and $w(x)$ the Wannier function, one has
\begin{equation}U=g\int dx|w(x)|^4\end{equation} \begin{equation}V=g\int
dx|w(x)w(x+d)|^2.\end{equation} 

The Hamiltonian Eq. (\ref{eq:Hspin}) 
can be brought to a fermionic form with the help of 
Jordan-Wigner transformation: 
\begin{eqnarray}
\s_i^+&=&a_i^\dagger\prod_{j<i}\s_i^z, \nonumber\\
\s_i^-&=&\left(\prod_{j<i}\s_i^z\right)a_i,\nonumber\\
\s_i^z&=&2a_i^\dagger a_i-1,
\end{eqnarray}
where the $a_i$'s are fermionic operator.  
Substituting these expressions into the spin-chain Hamiltonian we
get:
\begin{equation}
H=J\sum _i (a_i^\dagger a_{i+1}+h.c.)+
4V\sum_i a_i^\dagger a_i a_{i+1}^\dagger a_{i+1}.
\end{equation}
Thus we have obtained a 1D system of interacting spin polarized fermions
starting with a system of bosons.
Bosonizing the JW fermions we arrive at a LL Hamiltonian similar
to Eq. (\ref{eq:HHaldane}).

We should also mention that, recently, was shown that it is possible to study 
typical fermionic correlation phenomena using 2-level bosonic atoms trapped in 
optical lattices \cite{BelenCirac2002}. 

\section{Conclusion}\label{sec:conc}

In conclusion, we have described 1D (Fermi) gases under different trapping configurations 
as Luttinger liquids by using the availble exact solutions.
We have shown how it is possible to gain some physical insights with this approach also in the
non-interactig case (strictly speaking we cannot talk in this case of Luttinger liquids).
In addition, we have proposed some experiments to test the validity of this approach and to see in a clear
way the so-called ``spin-charge separation''. 

We emphasize that a series of approximations has been made. 
We considered our system as an infinite one, or better finite with periodic boundary conditions, 
to extract the Luttinger parameters. 
We commented on that for what concern an actual experiment.
The finite-size correction can be, in any case with some effort, taken into account.   
The analysis of the inhomogeneous case was carried out within a local density approximation. For
what we gave the range of validity.

In summary, cold quantum degenerate Fermi gases are a promising avenue to observe Luttinger physics
in a novel system in the laboratory.

\acknowledgments

Discussions with J.I. Cirac, J. von Delft, D. Jaksch and U.
Schollw\"ock are gratefully acknowledged. Work at Innsbruck
supported in part by the A.S.F., EU
Networks, and the Institute for Quantum Information. W.Z. and P.Z.
thank ITP, UCSB, for hospitality. P.Z. thanks also the MPQ, Garching, for
hospitality and the Alexander von Humboldt Award for support.
A.R. has been supported supported by the European 
Commission Project Cold Quantum Gases RTN Network Contract No. HPRN-CT-2000-00125.

\appendix

\section{Bosonization identities}

It is possible to represent the single fermion field operators
$\psi_{r,\s}$ ($r=R,L$ and $\s=\u,\d$) 
in terms of the boson field $\phi_\nu$ and $\Pi_\nu$ ($\nu = c,s$)
(see for instance \cite{Schulz:Review,vonDelft,Tsvelik,senechal}) by the bosonization identity
\begin{equation}
\psi_{r,\s}=\lim _{\alpha \rightarrow 0}\frac{F_{r,\s}}{(2\pi\alpha )^{1/2}}
e^{-a_r i2\sqrt{\pi }\phi _{R(L),\s}},
\label{eq:BI}
\end{equation}
where $a_{R(L)}=+(-)$. 
The fields $\phi_{r,\s}$ are recovered from Eq. (\ref{eq:bosonfield}).
The (hermitian) operators $F_{r,\s}$ are called Klein factors and obey the
Clifford algebra $\{F_{r,\s},F_{r',\s '}\}=2\delta_{r,r'}\delta_{\s, \s '}$. 
We remind that the Luttinger model is based on an infinite filled Dirac sea.
The last correspondence must be handled with utmost care when a product
of two (or more) fermion operators with the same argument are considered.
In such cases one must implement the normal ordering prescription,
i.e., $:A:=A-\langle A\rangle _{\rm vac} $.  For instance the right
way to compute the densities is by using the limit prescription: 
$ \rho(x)=\lim _{\varepsilon \rightarrow 0}:\psi ^{\dagger }(x)\psi 
(x+\varepsilon ):$ \cite{senechal}.\\
 It is useful to write the boson fields in terms of boson creation
and annihilation operators, $b_{\nu ,k}^\dagger$ and $ b_{\nu ,k}$,
$\nu=c,s$
\begin{eqnarray}
\phi _{\nu }(x) & = & \sqrt{\frac{K_{\nu }}{2L}}\sum _{k}
\frac{1}{\sqrt{|k|}}\left( b_{\nu ,k}e^{ikx}+b_{\nu ,k}^{\dagger
}e^{-ikx}\right) \nonumber \\
&+&\frac{\sqrt{\pi}x}{L}\hat{N}_\nu
\label{phi2q}
\end{eqnarray}
\begin{eqnarray}
\Pi _{\nu }(x) & = & i\sqrt{\frac{1}{2LK_{\nu }}}\sum _{k}
\sqrt{|k|}\left( b_{\nu ,k}e^{ikx}-b_{\nu ,k}^{\dagger }e^{-ikx}\right)
\nonumber \\ 
&-&\frac{\sqrt{\pi}}{L}\hat{J}_\nu.
\label{Pi2q}
\end{eqnarray}
The eigenvalues of the operators $\hat{N}_\nu$'s are 
the number of particles added to the ground state ($\nu=c$) and the
``magnetization''  ($N_\u-N_\d$) of the system with respect to the ground state ($\nu=s$). 
The operators $\hat{J}_\nu$ represent the respective currents. In the
thermodynamic limit ($L\rightarrow\infty$) or for excitations which do not
change neither the number of particles and the currents these operators can be
neglected. Note that such ``topological'' excitations can be recognized also 
in the exact solutions of the models we discussed in this paper. 

\subsection{Backward and umklapp scatterng}\label{app}

Using the previous identities it is possible to express also the backward 
scattering in terms of the bosonic fields $\phi_\nu$ and $\Pi_\nu$.
In terms of fermionic operators the backward scattering is 
\begin{equation}
H_b=g_1\sum_{k_1,k_2,\s} 
c_{R,\s}^\dagger(k_1)c_{L,\s}(k_1-q)c_{L,-\s}^\dagger(k_2)c_{R,-\s}(k_2+q)].
\end{equation}
Expressing the previous term in the real space and substituting the fermionic
fields with the Eq. (\ref{eq:BI}) one gets
\begin{equation}
{\mathcal H}_b(x)=4g_{1\perp}F_{R\u}F_{R\d}F_{L\u}F_{L\d}\cos(\sqrt{8}\phi_s).
\label{eq:Hback}
\end{equation}
The product of Klein factors can in this case substitute (with a kind of gauge
choice) with $+1$ (see \cite{Schulz:Review,senechal,vonDelft} for a comprehensive 
discussion). From the obtained expression is quite clear that, when $g_{1\perp}$
renormalizes to strong coupling, one has a gap for the spin excitations
\cite{emeryspin}. 
This is the case, e.g., for $g_{1\perp} < 0$ and $K_s=1$. 
When the lattice is present also an umklapp term can be present. 
The ``bosonized'' form of such term is essentially like Eq. (\ref{eq:Hback}), which 
$\phi_c$, instead of $\phi_s$ in the argument of the cosine function. But in this
case a gap can be opened only at half filling \cite{emerycharge}.

\end{document}